# Color Graphs: An Efficient Model For Two-Dimensional Cellular Automata Linear Rules

[1] BIRENDRA KUMAR NAYAK, [2] SUDHAKAR SAHOO
[3] SUSHANT KUMAR ROUT

[1] P.G.Department of Mathematics, Utkal University, Bhubaneswar-751004
bknatuu@yahoo.co.uk
[2] Department of CSEA, SIT, Silicon Hills, Patia, Bhubaneswar-751024
sudhakar.sahoo@gmail.com
[3] Department of Mathematics, Rayagada College, Rayagada, Orissa

*Abstract:* -. Two-dimensional nine neighbor hood rectangular Cellular Automata rules can be modeled using many different techniques like Rule matrices, State Transition Diagrams, Boolean functions, Algebraic Normal Form etc. In this paper, a new model is introduced using color graphs to model all the 512 linear rules. The graph theoretic properties therefore studied in this paper simplifies the analysis of all linear rules in comparison with other ways of its study.

## 1 Introduction

Originally John Von Neumann [3] introduced Cellular Automata (CA). A CA is a processing algorithm applied to a matrix of a fixed dimension with respect to space of locally connected cells whose mutual interactions determine their behaviors: time and space are discretized into intervals and cells, respectively. In forties mathematician Stanislas Ulam was interested in the evolution of graphics constructions generated by simple rules. The base of his constructions was a two-dimensional space divided into "cells", a sort of grid. Each of these cells could have two states: ON or OFF. Starting from a given pattern, the following generation was determined according to neighborhood rules. For example, if a cell was in contact with two "ON" cells, it would switch on to ON states; otherwise it would switch off. Ulam, who used one of the first computers, quickly noticed that this mechanism permitted to generate complex and graceful figures and that these figures could, in some cases, self-reproduce. Extremely simple rules permitted to build very complex patterns.

In this paper a 2-dimensional Cellular Automata is considered and each cell is influenced by its 9-neighboring cells including itself. The cells are characterized by its cell states; here we are considering the states are either 0 or 1. A 2-dimensional grid can be divided into several cells possessing states in binary. These binary data gives a binary information matrix (also referred as binary problem matrix or binary image) whose order is equal to the size of the grid say (m x n). Then, the rules of influences of each cell by its neighboring cells can be obtained by applying the rule to this binary information matrix, getting another binary information matrix, which gives the information about its next generation.

These influences can be captured through matrices [1, 8, 9], which we call as rule matrices. The binary information matrix of order (m x n) is reduced to a column matrix of order (mn x 1), arranged in a row major order, such that a (mn x mn) rule matrix must be necessary for applying it to this binary information matrix arranged in one-dimension. To study the properties of these rule matrices of 0's and 1's even for a small value of m and n are quite difficult. Although some interesting pattern and characteristics of these matrices has reported in [9] still there is a necessity to get a better handle to get some more information of these rules. In this paper, rule matrices are characterized by color graphs, which determine several interesting characteristics of rules.

In chapter 2, a review of earlier works in Cellular Automata and some ideas, which will be needed in our work, has been made. Here, we undertake the study of 2-dimesional CA where in each cell is assumed to be influenced by its 9 neighbors. The rules of influence are captured by the matrix representations of the linear Boolean functions with 9-variables. In chapter-3, we have studied the problems of 512 linear Boolean rules with 9-variables and the matrix representing them. In chapter-4, an attempt is being made to understand the graphical representation of rule matrices.

## 2 Review of Earlier work on 2-D CA

In 2-D Nine Neighborhood CA the next state of a particular cell is affected by the current state of itself and eight cells in its nearest neighborhood. Such dependencies are accounted by various rules. For the sake of simplicity, in this section we take into consideration only the linear rules, i.e. the rules, which can be realized by EX-OR operation only.

We want to find some algebraic properties of these $2^9$=512 linear CA rules that are quite useful to describe the behavior of 2-D Cellular Automata and its enumerable number of application reported in [2, 3, 6].
A specific rule convention that is adopted to number all 512 rules reported in [8, 9] is as follows:

| 64 | 128 | 256 |
|----|-----|-----|
| 32 | 1   | 2   |
| 16 | 8   | 4   |

Fig-1

The central box represents the current cell (i.e. the cell being considered) and all other boxes represent the eight nearest neighbors of that cell. The number within each box represents the rule number characterizing the dependency of the current cell on that particular neighbor only. Rule 1 characterizes dependency of the central cell on itself alone whereas such dependency only on its top neighbor is characterized by rule 128, and so on. These nine rules are called fundamental rules. In case the cell has dependency on two or more neighboring cells, the rule number will be the arithmetic sum of the numbers of the relevant cells. For example, the 2D CA rule 171 (128+32+8+2+1) refers to the five-neighborhood dependency of the central cell on (top, left, bottom, right and itself). The number of such rules is $^9C_0 + ^9C_1 + \ldots + ^9C_9$ =512 which includes rule characterizing no dependency.

The application of linear rules mentioned in the previous section can be realized on a problem matrix, where every entry is either 0 or 1.

**Illustration**
In figure 2, Rule 170 (2 + 8 + 32 + 128) is applied uniformly to each cell of a problem matrix of order (3 x 4) with null boundary condition (extreme cells are connected with logic-0 states).

$$\begin{pmatrix} 0 & 0 & 1 & 0 \\ 1 & 1 & 1 & 0 \\ 1 & 0 & 1 & 1 \end{pmatrix} \xrightarrow{Rule\ 170} \begin{pmatrix} 1 & 0 & 1 & 1 \\ 0 & 0 & 1 & 0 \\ 1 & 1 & 0 & 1 \end{pmatrix}$$

[Fig-2: shows the cell under consideration, will change its state by adding the states of its 4 orthogonal neighboring cells. 0+1+1+1=1. Similarly, Rule 170 is applied to all other cells.]

Also for this problem matrix of dimension (3 x 4), Rule 170 using an algorithm reported in [9, 10] can be realized as another binary matrix of dimension (12 x 12) called rule matrix for rule 170 which when multiplied with a (12 x 1) column vector (in this case [0 0 1 0 1 1 1 0 1 0 1 1]$^T$ ) formed from the problem matrix in a row major order gives the same output image.

# 3 Graph representation of Basic linear rules

As rule matrices are basically binary square matrices of dimension (mn x mn) for a binary information matrix (m x n), therefore these matrices can be realized as adjacency matrices that give a graph with vertices and edges. As every adjacency matrix is a vertex-vertex relationship so each row (or column) of the Rule matrix corresponds to a vertex of the graph, which also refers to as each cell of the binary information matrix. The position of 1's present in each row in the Rule matrix represents the nine-neighbor hood dependency of the cell for that row. So for a particular (m x n) binary image one can construct 512 different graphs for 512 rules. All these graphs contain mn number of vertices numbered as $v_1, v_2, \ldots v_{mn}$ where vertex $v_i$ corresponds to $i^{th}$ cell counting starting from the left most corner (0, 0) position of (m x n) problem matrix in a row major order and the edges between the vertices denotes their nine-neighborhood dependency for their corresponding cell positions in the problem matrix.

There are five fundamental rules Rule1, Rule2, Rule4, Rule 8 and Rule 16 shown in Fig-1 are known as basic rules also discussed in [10] as other fundamental rules Rule32, Rule64, Rule128 and Rule 256 can be obtained from these with a transpose operation. The graphical representation of these 5 basic rules and other fundamental rules are explained through some theorems as follows:

**Theorem 3.1:** For an m x n problem matrix, the rule graph for Rule 1(i.e. $G_1$) has mn self loops.

**Example 3.1**

Let us consider a 2 x 3 problem matrix, it's rule matrix is of order (6 x 6) (i.e. $M_{6x6}$)
For Rule 1 (i.e. $M_1$), the matrix form is

$$M_1 = \begin{pmatrix} 1 & 0 & 0 & 0 & 0 & 0 \\ 0 & 1 & 0 & 0 & 0 & 0 \\ 0 & 0 & 1 & 0 & 0 & 0 \\ 0 & 0 & 0 & 1 & 0 & 0 \\ 0 & 0 & 0 & 0 & 1 & 0 \\ 0 & 0 & 0 & 0 & 0 & 1 \end{pmatrix}$$

*It's graphical form is*

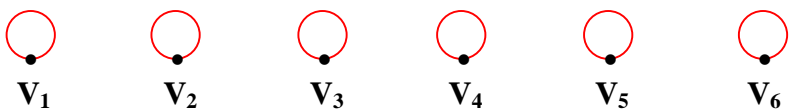

$V_1 \quad V_2 \quad V_3 \quad V_4 \quad V_5 \quad V_6$

[The graph has 6 independent loops]

**Theorem 3.2:** For an m x n problem matrix, the rule matrix for Rule 2 (i.e. $M_2$) has m components and each component has n-vertices. Therefore vertex $V_i$ of Rule 2 graph connects with vertex $V_{i+1}$ except i = kn, n ∈ N.

**Example 3.2**

Let us consider a 2x3 problem matrix, it's rule matrix is of order 6 x 6(i.e. $M_{6x6}$).
For Rule 2(i.e. $M_2$), the matrix form is

$$M_2 = \begin{pmatrix} 0 & 1 & 0 & 0 & 0 & 0 \\ 0 & 0 & 1 & 0 & 0 & 0 \\ 0 & 0 & 0 & 0 & 0 & 0 \\ 0 & 0 & 0 & 0 & 1 & 0 \\ 0 & 0 & 0 & 0 & 0 & 1 \\ 0 & 0 & 0 & 0 & 0 & 0 \end{pmatrix}$$

*It's graphical form is*

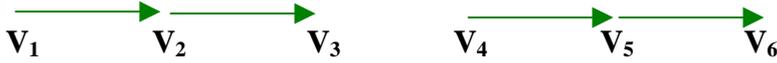

[The graph has two components and each component has 3 vertices]

**Theorem 3.3:** For an (m x n) problem matrix, the vertex $V_i$ of Rule 4 graph connects with vertex $V_{n+i+1}$ for i=1,2,…mn - n-1 except i = n and mn – (n-1) i.e. $V_n$ and $V_{mn-n+1}$ are two isolated vertices.

**Example 3.3**

Let us consider a 3x4 problem matrix, it's rule matrix is of order 12 x 12(i.e. $M_{12x12}$).

For Rule 4(i.e. $M_4$), the matrix form is

$$M_4 = \begin{pmatrix} 0 & 0 & 0 & 0 & 0 & 0 & 0 & 0 & 0 & 0 & 0 & 0 \\ 0 & 0 & 0 & 0 & 0 & 0 & 0 & 0 & 0 & 0 & 0 & 0 \\ 0 & 0 & 0 & 0 & 0 & 1 & 0 & 0 & 0 & 0 & 0 & 0 \\ 0 & 0 & 0 & 0 & 0 & 0 & 1 & 0 & 0 & 0 & 0 & 0 \\ 0 & 0 & 0 & 0 & 0 & 0 & 0 & 1 & 0 & 0 & 0 & 0 \\ 0 & 0 & 0 & 0 & 0 & 0 & 0 & 0 & 0 & 0 & 0 & 0 \\ 0 & 0 & 0 & 0 & 0 & 0 & 0 & 0 & 0 & 1 & 0 & 0 \\ 0 & 0 & 0 & 0 & 0 & 0 & 0 & 0 & 0 & 0 & 1 & 0 \\ 0 & 0 & 0 & 0 & 0 & 0 & 0 & 0 & 0 & 0 & 0 & 1 \\ 0 & 0 & 0 & 0 & 0 & 0 & 0 & 0 & 0 & 0 & 0 & 0 \\ 0 & 0 & 0 & 0 & 0 & 0 & 0 & 0 & 0 & 0 & 0 & 0 \\ 0 & 0 & 0 & 0 & 0 & 0 & 0 & 0 & 0 & 0 & 0 & 0 \end{pmatrix}$$

*It's graphical form is*

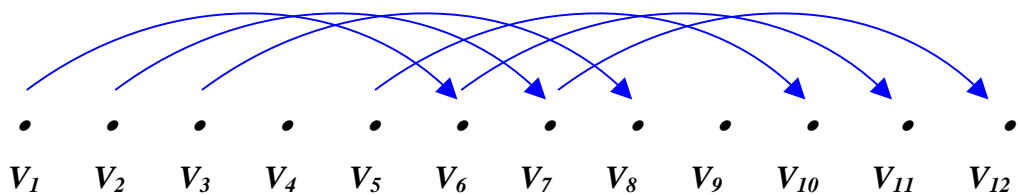

Here $V_5$ and $V_9$ are isolated vertices.

**Theorem 3.4:** For an m x n problem matrix, the vertex $V_i$ of Rule 8 graph connects to vertex $V_{n+i}$, that is the edges are $(V_i, V_{n+i})$ for i=1,2,…,n.

**Example 3.4** Let us consider a 2x4 problem matrix, its rule matrix is of order 8 x 8(i.e. $M_{8x8}$).

For Rule 8(i.e. $M_8$), the matrix form is

$$M_8 = \begin{pmatrix} 0 & 0 & 0 & 0 & 1 & 0 & 0 & 0 \\ 0 & 0 & 0 & 0 & 0 & 1 & 0 & 0 \\ 0 & 0 & 0 & 0 & 0 & 0 & 1 & 0 \\ 0 & 0 & 0 & 0 & 0 & 0 & 0 & 1 \\ 0 & 0 & 0 & 0 & 0 & 0 & 0 & 0 \\ 0 & 0 & 0 & 0 & 0 & 0 & 0 & 0 \\ 0 & 0 & 0 & 0 & 0 & 0 & 0 & 0 \\ 0 & 0 & 0 & 0 & 0 & 0 & 0 & 0 \end{pmatrix}$$

*It's graphical form is*

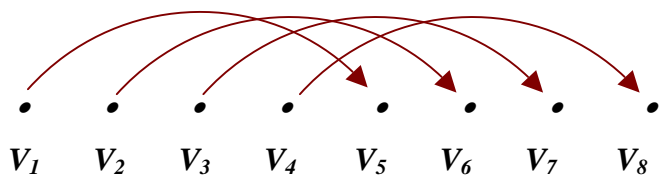

$V_1 \quad V_2 \quad V_3 \quad V_4 \quad V_5 \quad V_6 \quad V_7 \quad V_8$

**Theorem 3.5:** For an m x n problem matrix, the vertices $V_1$ and $V_{mn}$ of Rule 16 graph are isolated, other pair of vertices are joined in the sequence $(V_i, V_{n+i-1})$ for i=2 to (mn-2).

**Example 3.5**

Let us consider a 2 x 3 problem matrix, it's rule matrix is of order 6 x 6 (i.e. $M_{6x6}$).

For Rule 16(i.e. $M_{16}$), the matrix form is

$$M_{16} = \begin{pmatrix} 0 & 0 & 0 & 0 & 0 & 0 \\ 0 & 0 & 0 & 1 & 0 & 0 \\ 0 & 0 & 0 & 0 & 1 & 0 \\ 0 & 0 & 0 & 0 & 0 & 0 \\ 0 & 0 & 0 & 0 & 0 & 0 \\ 0 & 0 & 0 & 0 & 0 & 0 \end{pmatrix}$$

*Its graphical form is*

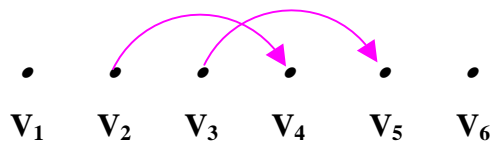

$V_1 \quad V_2 \quad V_3 \quad V_4 \quad V_5 \quad V_6$

Here $V_1$ and $V_6$ are isolated vertices.

**Theorem 3.6:** The graph for Rule 32, Rule 64, Rule 128 and Rule 256 are same as Rule 2, Rule 4, Rule 8 and Rule16 respectively with opposite direction as corresponding matrices are transpose to each other reported in [9].

**Example 3.6**

Let us consider a (2 x 3) problem matrix, the rule matrix is of order (6 x 6)

$$M_2 = \begin{pmatrix} 0 & 1 & 0 & 0 & 0 & 0 \\ 0 & 0 & 1 & 0 & 0 & 0 \\ 0 & 0 & 0 & 0 & 0 & 0 \\ 0 & 0 & 0 & 0 & 1 & 0 \\ 0 & 0 & 0 & 0 & 0 & 1 \\ 0 & 0 & 0 & 0 & 0 & 0 \end{pmatrix} \quad M_{32} = M_2^T = \begin{pmatrix} 0 & 0 & 0 & 0 & 0 & 0 \\ 1 & 0 & 0 & 0 & 0 & 0 \\ 0 & 1 & 0 & 0 & 0 & 0 \\ 0 & 0 & 0 & 0 & 0 & 0 \\ 0 & 0 & 0 & 1 & 0 & 0 \\ 0 & 0 & 0 & 0 & 1 & 0 \end{pmatrix}$$

**Rule 2 graph**

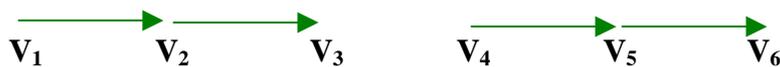

**Rule 32 graph**

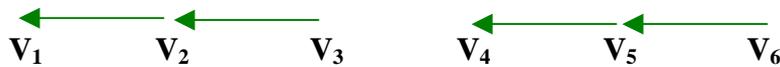

# 4 Construction of graphs of all linear rules from the basic graphs

In this section a new operation called join operation called Join operation is introduced which is used to combine some or all of the basic graphs to form a new graph.

**Definition 4.1:** If there is edge between two vertices, it is considered as 1 and if there is no edge between two vertices, then it is considered as 0.

The join operation is defined as follows

$$0 + 0 = 0$$
$$1 + 0 = 1$$
$$0 + 1 = 1$$
$$1 + 1 = 0$$

When two edges between two vertices are joined the result is no edge between them.

**Example 4.1**

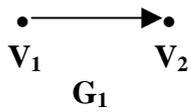 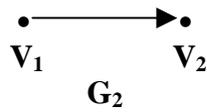

$G_1$      $G_2$

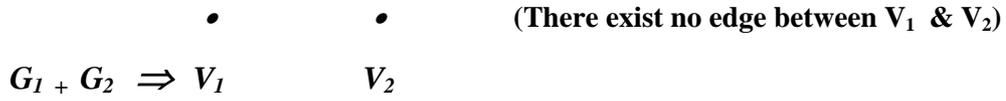

$G_1 + G_2 \Rightarrow V_1 \quad V_2$      (There exist no edge between $V_1$ & $V_2$)

**Example 4.2**

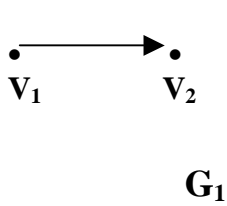      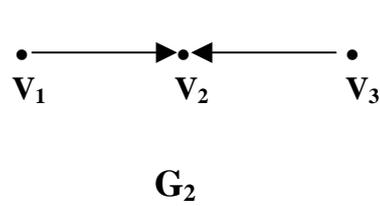

$G_1$      $G_2$

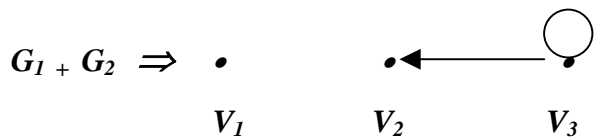

$G_1 + G_2 \Rightarrow$

$V_1 \quad V_2 \quad V_3$

**Example 4.3** Consider a (2 x 2) problem matrix. Its Rule Matrix is of order (4 x 4)

The graph of Rule 1 that is $G_1$ corresponding to matrix $M_1$ is as follows

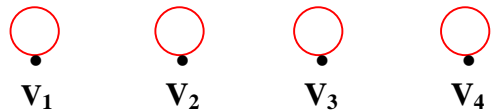

The graph of Rule 2 i.e. $G_2$ corresponding to matrix $M_2$ is

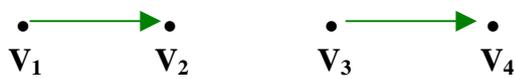

The graph of Rule 4 i.e. $M_4$ is

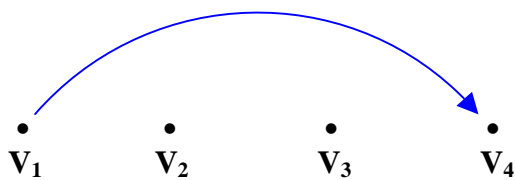

The graph of Rule 8 i.e. $M_8$ is

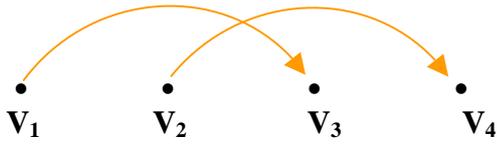

The graph of Rule 16 i.e. $M_{16}$ is

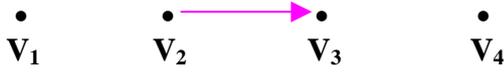

Let us find the graph of Rule 7 i.e. $G_7$ corresponding to the rule matrix $M_7 = M_1 + M_2 + M_4$

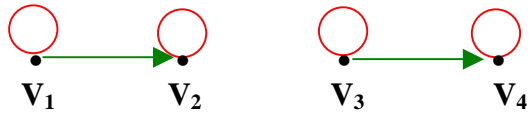

The graph is obtained by the join operation of the above three graphs Rule 1, Rule 2, and Rule 4. Similarly other graphs can be obtained by the join operation of five basic Rule graphs, some of them are shown below.

**MATRIX**            **GRAPH**

1) $M_1 = \begin{bmatrix} 1 & 0 & 0 & 0 \\ 0 & 1 & 0 & 0 \\ 0 & 0 & 1 & 0 \\ 0 & 0 & 0 & 1 \end{bmatrix}$

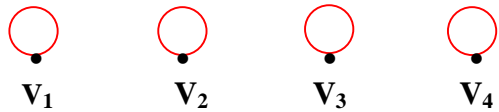

**Undirected, looped**

2) $M_2 = \begin{bmatrix} 0 & 1 & 0 & 0 \\ 0 & 0 & 0 & 0 \\ 0 & 0 & 0 & 1 \\ 0 & 0 & 0 & 0 \end{bmatrix}$

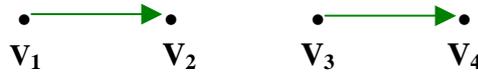

**Directed, simple**

3) $M_3 = \begin{bmatrix} 1 & 1 & 0 & 0 \\ 0 & 1 & 0 & 0 \\ 0 & 0 & 1 & 1 \\ 0 & 0 & 0 & 1 \end{bmatrix}$

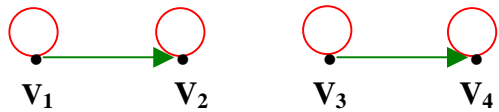

**Directed, looped**

4) $M_4 = \begin{bmatrix} 0 & 0 & 0 & 1 \\ 0 & 0 & 0 & 0 \\ 0 & 0 & 0 & 0 \\ 0 & 0 & 0 & 0 \end{bmatrix}$

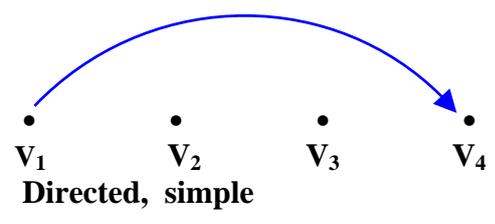

5) $M_5 = \begin{bmatrix} 1 & 0 & 0 & 1 \\ 0 & 1 & 0 & 0 \\ 0 & 0 & 1 & 0 \\ 0 & 0 & 0 & 1 \end{bmatrix}$

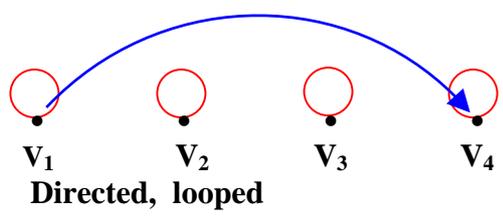

6) $M_6 = \begin{bmatrix} 0 & 1 & 0 & 1 \\ 0 & 0 & 0 & 0 \\ 0 & 0 & 0 & 1 \\ 0 & 0 & 0 & 0 \end{bmatrix}$

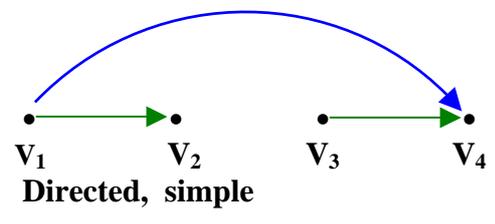

7) $M_9 = \begin{bmatrix} 1 & 0 & 1 & 0 \\ 0 & 1 & 0 & 1 \\ 0 & 0 & 1 & 0 \\ 0 & 0 & 0 & 1 \end{bmatrix}$

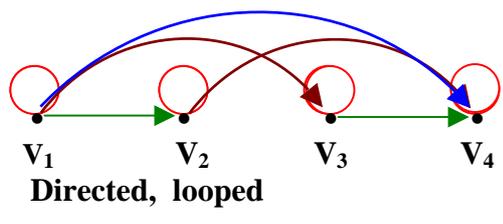

8) $M_{10} = \begin{bmatrix} 0 & 1 & 1 & 0 \\ 0 & 0 & 0 & 1 \\ 0 & 0 & 0 & 1 \\ 0 & 0 & 0 & 0 \end{bmatrix}$

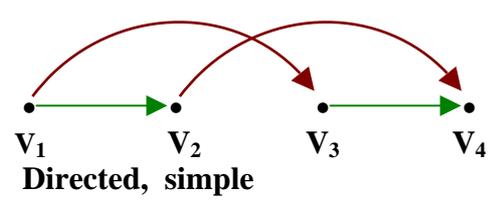

9) $M_{11} = \begin{bmatrix} 1 & 1 & 1 & 0 \\ 0 & 1 & 0 & 1 \\ 0 & 0 & 1 & 1 \\ 0 & 0 & 0 & 1 \end{bmatrix}$

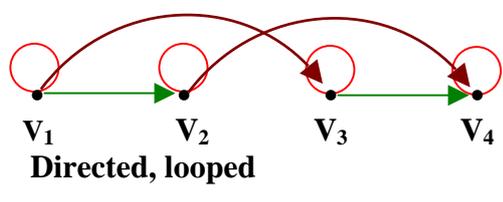

10) $M_{12} = \begin{bmatrix} 0 & 0 & 1 & 1 \\ 0 & 0 & 0 & 1 \\ 0 & 0 & 0 & 0 \\ 0 & 0 & 0 & 0 \end{bmatrix}$

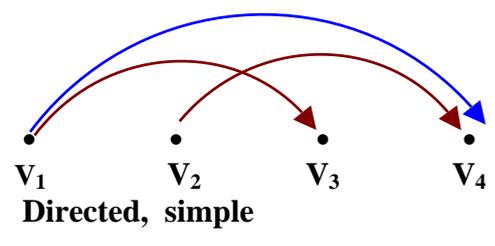

11) $M_{13} = \begin{bmatrix} 1 & 0 & 1 & 1 \\ 0 & 1 & 0 & 1 \\ 0 & 0 & 1 & 0 \\ 0 & 0 & 0 & 1 \end{bmatrix}$

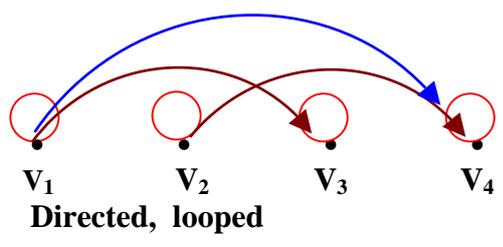

$V_1$    $V_2$    $V_3$    $V_4$
**Directed, looped**

12) $M_{14} = \begin{bmatrix} 0 & 1 & 1 & 1 \\ 0 & 0 & 0 & 1 \\ 0 & 0 & 0 & 1 \\ 0 & 0 & 0 & 0 \end{bmatrix}$

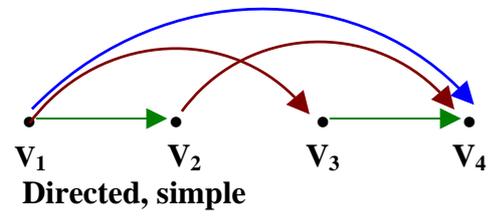

$V_1$    $V_2$    $V_3$    $V_4$
**Directed, simple**

13) $M_{15} = \begin{bmatrix} 1 & 1 & 1 & 1 \\ 0 & 1 & 0 & 1 \\ 0 & 0 & 1 & 1 \\ 0 & 0 & 0 & 1 \end{bmatrix}$

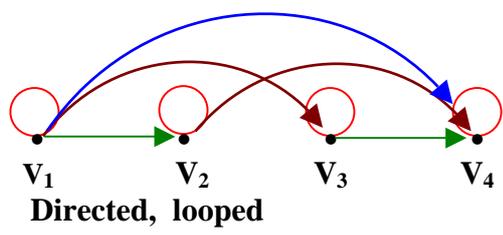

$V_1$    $V_2$    $V_3$    $V_4$
**Directed, looped**

14) $M_{21} = \begin{bmatrix} 1 & 0 & 0 & 1 \\ 0 & 1 & 1 & 0 \\ 0 & 0 & 1 & 0 \\ 0 & 0 & 0 & 1 \end{bmatrix}$

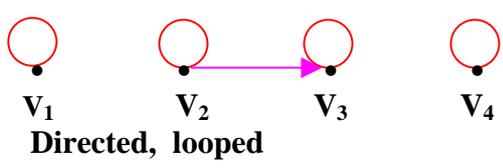

$V_1$    $V_2$    $V_3$    $V_4$
**Directed, looped**

15) $M_{22} = \begin{bmatrix} 0 & 1 & 0 & 1 \\ 0 & 0 & 1 & 0 \\ 0 & 0 & 0 & 1 \\ 0 & 0 & 0 & 0 \end{bmatrix}$

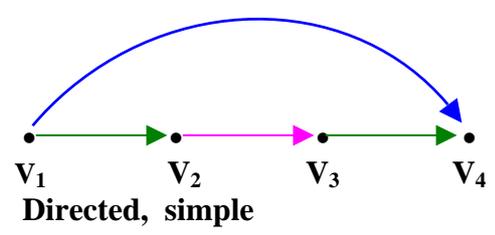

$V_1$    $V_2$    $V_3$    $V_4$
**Directed, simple**

16) $M_{25} = \begin{bmatrix} 1 & 0 & 1 & 0 \\ 0 & 1 & 1 & 1 \\ 0 & 0 & 1 & 0 \\ 0 & 0 & 0 & 1 \end{bmatrix}$

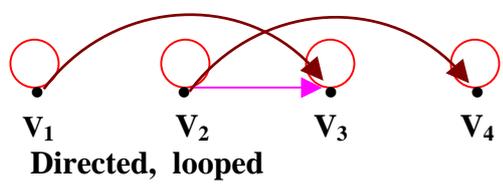

$V_1$    $V_2$    $V_3$    $V_4$
**Directed, looped**

17) $M_{104} = \begin{bmatrix} 0 & 0 & 1 & 0 \\ 1 & 0 & 0 & 1 \\ 0 & 0 & 0 & 0 \\ 1 & 0 & 1 & 0 \end{bmatrix}$

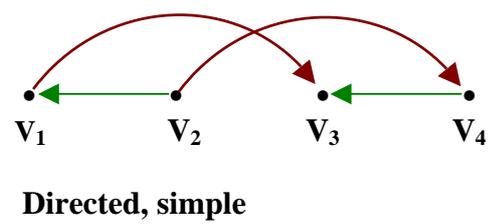

$V_1$    $V_2$    $V_3$    $V_4$

**Directed, simple**

18) $M_{105} = \begin{bmatrix} 1 & 0 & 1 & 0 \\ 1 & 1 & 0 & 1 \\ 0 & 0 & 1 & 0 \\ 1 & 0 & 1 & 1 \end{bmatrix}$ 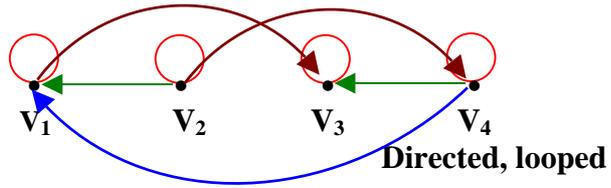

**Directed, looped**

19) $M_{116} = \begin{bmatrix} 0 & 0 & 0 & 1 \\ 1 & 0 & 1 & 0 \\ 0 & 0 & 0 & 0 \\ 1 & 0 & 1 & 0 \end{bmatrix}$ 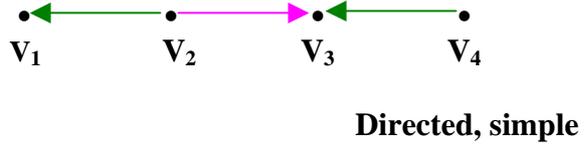

**Directed, simple**

20) $M_{117} = \begin{bmatrix} 1 & 0 & 0 & 1 \\ 1 & 1 & 1 & 0 \\ 0 & 0 & 1 & 0 \\ 1 & 0 & 1 & 1 \end{bmatrix}$ 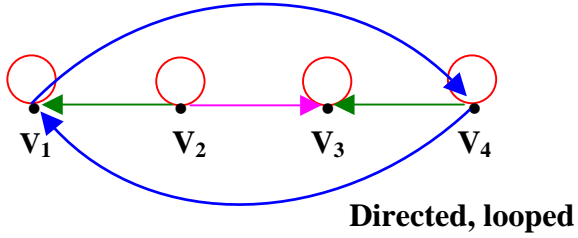

**Directed, looped**

21) $M_{119} = \begin{bmatrix} 1 & 1 & 0 & 1 \\ 1 & 1 & 1 & 0 \\ 0 & 0 & 1 & 1 \\ 1 & 0 & 1 & 1 \end{bmatrix}$ 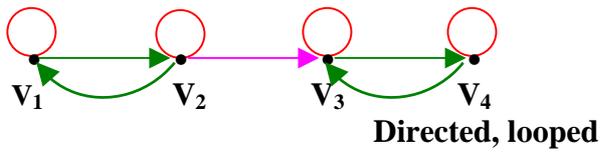

**Directed, looped**

22) $M_{121} = \begin{bmatrix} 1 & 0 & 1 & 0 \\ 1 & 1 & 1 & 1 \\ 0 & 0 & 1 & 0 \\ 1 & 0 & 1 & 1 \end{bmatrix}$ 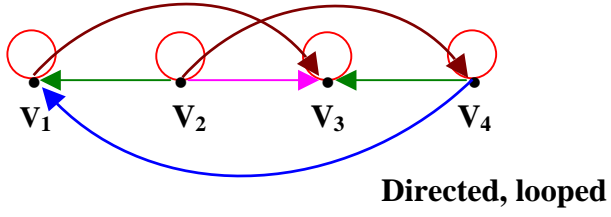

**Directed, looped**

23) $M_{122} = \begin{bmatrix} 0 & 1 & 1 & 0 \\ 1 & 0 & 1 & 1 \\ 0 & 0 & 0 & 1 \\ 1 & 0 & 1 & 0 \end{bmatrix}$ 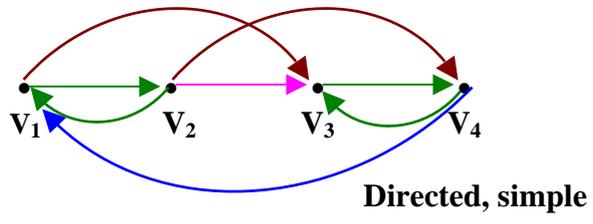

**Directed, simple**

24) $M_{128} = \begin{bmatrix} 0 & 0 & 0 & 0 \\ 0 & 0 & 0 & 0 \\ 1 & 0 & 0 & 0 \\ 0 & 1 & 0 & 0 \end{bmatrix}$ 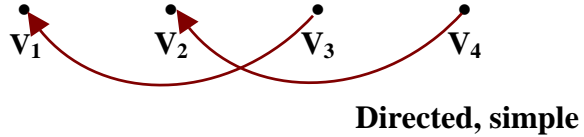

**Directed, simple**

25) $M_{129} = \begin{bmatrix} 1 & 0 & 0 & 0 \\ 0 & 1 & 0 & 0 \\ 1 & 0 & 1 & 0 \\ 0 & 1 & 0 & 1 \end{bmatrix}$

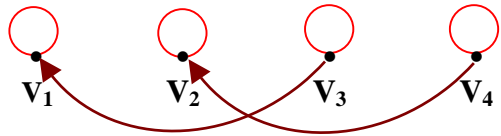

**Directed, looped**

26) $M_{290} = \begin{bmatrix} 0 & 1 & 0 & 0 \\ 1 & 0 & 0 & 0 \\ 0 & 1 & 0 & 0 \\ 0 & 0 & 1 & 0 \end{bmatrix}$

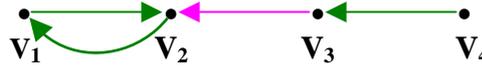

**Directed, simple**

27) $M_{297} = \begin{bmatrix} 1 & 0 & 1 & 0 \\ 1 & 1 & 0 & 1 \\ 0 & 1 & 1 & 0 \\ 0 & 0 & 1 & 1 \end{bmatrix}$

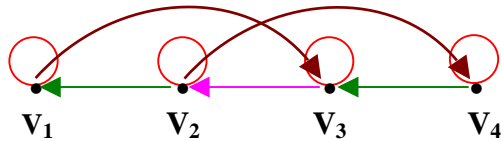

**Directed, looped**

28) $M_{299} = \begin{bmatrix} 1 & 1 & 1 & 0 \\ 1 & 1 & 0 & 1 \\ 0 & 1 & 1 & 1 \\ 0 & 0 & 1 & 1 \end{bmatrix}$

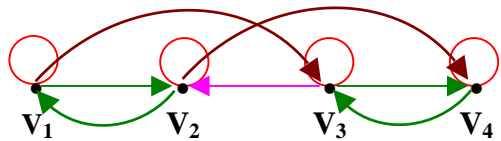

**Directed, looped**

29) $M_{301} = \begin{bmatrix} 1 & 0 & 1 & 1 \\ 1 & 1 & 0 & 1 \\ 0 & 1 & 1 & 0 \\ 0 & 0 & 1 & 1 \end{bmatrix}$

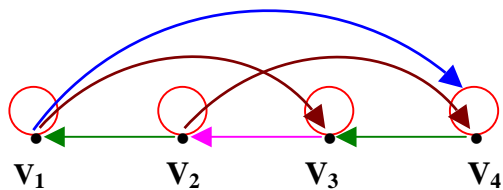

**Directed, looped**

30) $M_{354} = \begin{bmatrix} 0 & 1 & 0 & 0 \\ 1 & 0 & 0 & 0 \\ 0 & 1 & 0 & 1 \\ 1 & 0 & 1 & 0 \end{bmatrix}$

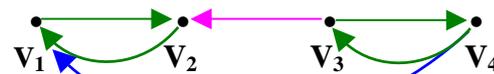

**Directed, simple**

31) $M_{463} = \begin{bmatrix} 1 & 1 & 1 & 1 \\ 0 & 1 & 0 & 1 \\ 1 & 1 & 1 & 1 \\ 1 & 1 & 0 & 1 \end{bmatrix}$

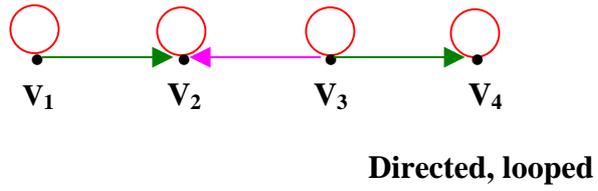

**Directed, looped**

32) $M_{466} = \begin{bmatrix} 0 & 1 & 0 & 0 \\ 0 & 0 & 1 & 0 \\ 1 & 1 & 0 & 1 \\ 1 & 1 & 0 & 0 \end{bmatrix}$

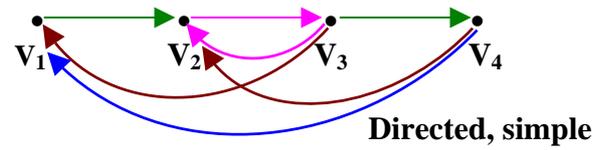

**Directed, simple**

33) $M_{472} = \begin{bmatrix} 0 & 0 & 1 & 0 \\ 0 & 0 & 1 & 1 \\ 1 & 1 & 0 & 0 \\ 1 & 1 & 0 & 0 \end{bmatrix}$

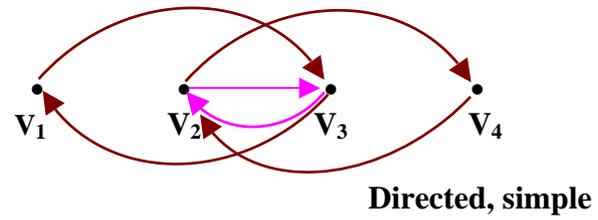

**Directed, simple**

34) $M_{474} = \begin{bmatrix} 0 & 1 & 1 & 0 \\ 0 & 0 & 1 & 1 \\ 1 & 1 & 0 & 1 \\ 1 & 1 & 0 & 0 \end{bmatrix}$

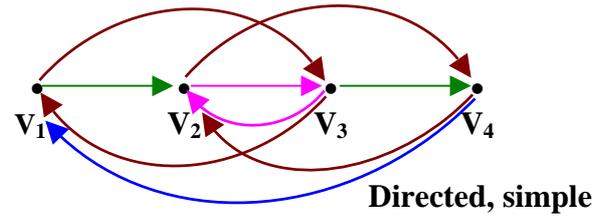

**Directed, simple**

35) $M_{501} = \begin{bmatrix} 1 & 0 & 0 & 1 \\ 1 & 1 & 1 & 0 \\ 1 & 1 & 1 & 0 \\ 1 & 1 & 1 & 1 \end{bmatrix}$

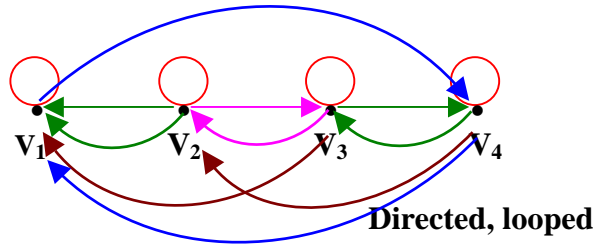

**Directed, looped**

36) $M_{505} = \begin{bmatrix} 1 & 0 & 1 & 0 \\ 1 & 1 & 1 & 1 \\ 1 & 1 & 1 & 0 \\ 1 & 1 & 1 & 1 \end{bmatrix}$

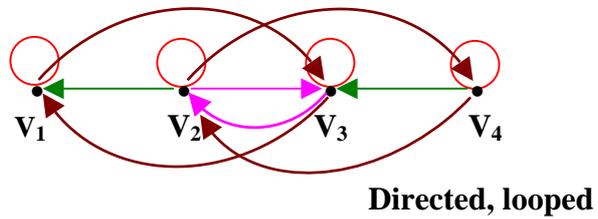

**Directed, looped**

37) $M_{509} = \begin{bmatrix} 1 & 0 & 1 & 1 \\ 1 & 1 & 1 & 1 \\ 1 & 1 & 1 & 0 \\ 1 & 1 & 1 & 1 \end{bmatrix}$

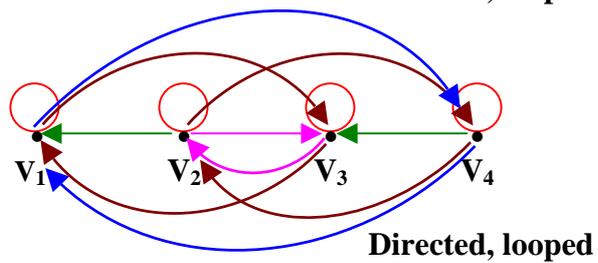

**Directed, looped**

38) $M_{510} = \begin{bmatrix} 0 & 1 & 1 & 1 \\ 1 & 0 & 1 & 1 \\ 1 & 1 & 0 & 1 \\ 1 & 1 & 1 & 0 \end{bmatrix}$

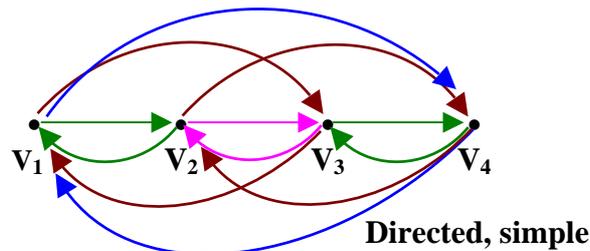

**Directed, simple**

39) $M_{511} = \begin{bmatrix} 1 & 1 & 1 & 1 \\ 1 & 1 & 1 & 1 \\ 1 & 1 & 1 & 1 \\ 1 & 1 & 1 & 1 \end{bmatrix}$

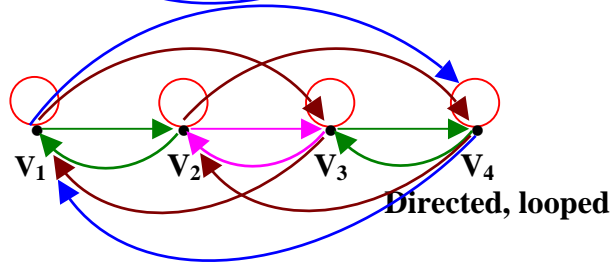

**Directed, looped**

## 6  Conclusion

Graph representation is always easy to visualize than a matrix representation with 0's and 1's. In this paper 512 linear rules are represented by directed graphs and as enumerable applications are there using graphs in computer science and other areas. Therefore we conjecture that all the fundamental graphs and other graphs can be used in the same application.